\documentstyle[12pt]{article}
\newlength{\mathspace}
\def\np#1{ Nucl. Phys. B#1}
\def\pr#1    { Phys. Rev. D#1 }
\def\pl#1{ Phys. Lett. B#1}

\def\ijmp#1  { Int. Jour. Mod. Phys. A#1 }
\def\mpl#1   { Mod. Phys. Lett. A#1 }
\def\begineq{\begin{equation}}
\def\endeq{\end{equation}}
\def\eqabegin{\begin{eqnarray}}
\def\eqaend{\end{eqnarray}}
\def\nn{\nonumber}

\begin{document}
\baselineskip=0.7cm
\setlength{\mathspace}{2.5mm}

%%%%%%%%%%%%%%%%%%%%%%%%%%%%%%%%%%%%%%%%%%%%%%%%%%%%%%%%%%

                                %titlepage

%%%%%%%%%%%%%%%%%%%%%%%%%%%%%%%%%%%%%%%%%%%%%%%%%%%%%%%%%%
\begin{titlepage}

    \begin{normalsize}
     \begin{flushright}
                 US-FT-48/96 \\
                 hep-th/9612141\\
     \end{flushright}
    \end{normalsize}
    \begin{LARGE}
       \vspace{1cm}
       \begin{center}
         {Orbifolds of M-Theory and}\\ 
         {Type II String Theories in Two Dimensions}\\ 
       \end{center}
    \end{LARGE}

  \vspace{5mm}

\begin{center}
           
             \vspace{.5cm}

            Shibaji R{\sc oy}
           \footnote{E-mail address:
              roy@gaes.usc.es}

                 \vspace{2mm}

        {\it Departamento de F\'\i sica de Part\'\i culas} \\
        {\it Universidade de Santiago,}\\
        {\it E-15706 Santiago de Compostela, Spain}\\
      \vspace{2.5cm}

    \begin{large} ABSTRACT \end{large}
        \par
\end{center}
 \begin{normalsize}
\ \ \ \
We consider several orbifold compactifications of M-theory and their 
corresponding type II duals in two space-time dimensions. In particular, we 
show that while the orbifold compactification of M-theory on 
T$^9$/${\cal J}_9$ is dual to the orbifold compactification of type IIB string 
theory on T$^8$/${\cal I}_8$, the same orbifold T$^8$/${\cal I}_8$ of type IIA 
string theory is dual to M-theory compactified on a smooth product manifold 
K3 $\times$ T$^5$. Similarly, while the orbifold compactification of M-theory 
on (K3 $\times$ T$^5$)/$\sigma \cdot {\cal J}_5$ is dual to the orbifold 
compactification of type IIB string theory on 
(K3 $\times$ T$^4$)/$\sigma \cdot {\cal I}_4$, the same orbifold of type IIA 
string theory is dual to the orbifold 
T$^4$ $\times$ (K3 $\times$ S$^1$)/$\sigma \cdot {\cal J}_1$ of M-theory. The 
spectrum of various orbifold compactifications of M-theory and type II string 
theories on both sides are compared giving evidence in favor of these duality 
conjectures. We also comment on a connection between Dasgupta-Mukhi-Witten 
conjecture and Dabholkar-Park-Sen conjecture for the six-dimensional orbifold 
models of type IIB string theory and M-theory.
\end{normalsize}

\end{titlepage}
\vfil\eject

\begin{large}
\noindent{\bf 1. Introduction:}
\end{large}

\vspace{.5cm}

During last one year or so a large number of dual pairs involving string 
theories [1--10], M-theory [11--13] and F-theory [14,15] have been constructed 
by assuming that the
orbifolding procedure commutes with duality. In this process, once a dual
pair is identified one can obtain another dual pair by further compactifying
the theories on an internal manifold ${\cal M}$ and then modding out the
original pairs by a combined group of discrete symmetries representing an
internal symmetry transformation of the original theories as well as any
geometric action ($s$) on the internal manifold. There are, however, notable 
examples where this procedure breaks down. In fact, all the different cases
of constructing dual pairs through orbifolding procedure have been classified
by Sen [9]. They fall into three categories depending on whether the geometric
action `$s$' acts freely (without fixed points) on the internal manifold
${\cal M}$ or not or whether it acts trivially on ${\cal M}$ and has been 
referred to as categories 1, 2 and 3 respectively. (There are some subtleties
involved in this procedure for more general orbifold group {\bf Z}$_N$ with 
$N>2$ [16--20].)
The cases where orbifolding procedure is known not to commute with duality
fall into category 3 [9]. It has been shown in [9] through a number of 
examples in string
theory, that this procedure of obtaining dual pairs works for the cases
1 and 2. One of the interesting aspects of this classification is that when the
orbifolding procedure falls into categories 1 and 2, it gives sensible dual 
pairs even when the original dual pairs involve M-theory (if we take M-theory
by definition as the strong coupling or the large radius limit of type IIA 
string theory) and was also emphasized in [12]. This is quite remarkable 
since this way one can connect various
compactifications of M-theory to the known compactifications of string theory
and extract certain properties of the former without knowing much about its
world volume theory.

In this paper, we study examples of dual pairs involving M-theory and type II 
string theories in two dimensions which can be obtained from the M-theory
definition\footnote[1]{It has been demonstrated beautifully in a recent
paper by Sen [21] that all the conjectured duality involving string theories,
M- and F-theory compactifications can be `derived' from a single 
non-perturbative duality between type I and heterotic string theory in ten
dimensions with the gauge group SO(32), T-dualities and the definition of M-
and F-theories.} (i.e. the equivalence of M-theory compactified on S$^1$ 
and type IIA string theory when the radius of the circle goes to zero) through
orbifolding procedure of category 2 in the classification made by Sen.
It has been conjectured by Dasgupta and Mukhi [22] and also by Sen [12]
that the two
dimensional orbifold model T$^9$/$\{1,\,{\cal J}_9\}$ of M-theory is
equivalent to the two dimensional orbifold model T$^8$/$\{1,\,{\cal I}_8\}$
of type IIB theory, where ${\cal J}_9$ changes the sign of the coordinates of
T$^9$ as well as the sign of the three-form gauge field present in M-theory,
whereas ${\cal I}_8$ simply changes the sign of the coordinates of T$^8$.
Equivalence of these two models follows from the M-theory definition and 
through the orbifolding procedure of category 2.
By comparing the massless spectrum on both sides,
we give further evidence in favor of this conjecture. We show that this
two dimensional type IIB model has a gravity multiplet and 8 scalar multiplets
as massless spectrum in the untwisted sector. In the twisted sector it has 256
antichiral bosons coming from the 256 fixed points of the orbifold and remain
inert under (0, 16) supersymmetry of the model. An identical spectrum has 
been obtained in the M-theory orbifold model on T$^9$/$\{1,\,{\cal J}_9\}$
in ref.[22]. It has been noted in ref.[17], that
the orbifold T$^8$/$\{1,\,{\cal I}_8\}$ of type IIB theory can not be smoothed
out to any known manifold unlike the orbifold T$^4$/$\{1,\,{\cal I}_4\}$ 
which is known to give a
smooth manifold K3. The situation is different for type IIA compactification
on the same orbifold, namely, T$^8$/$\{1,\,{\cal I}_8\}$. In this case, we show
the  equivalence between type IIA model on  T$^8$/$\{1,\,{\cal I}_8\}$
and K3 $\times$ T$^4$. In fact, this type IIA orbifold model is equivalent
to heterotic string theory on T$^8$, type I theory on T$^8$, type IIA (or type 
IIB) theory
on K3 $\times$ T$^4$, M-theory on T$^4$ $\times$ T$^5$/$\{1,\,{\cal J}_5\}$  
and M-theory on K3
$\times$ T$^5$. We thus find the equivalence of type IIA theory compactified
on two completely different internal spaces namely, one is an orbifold and
the other is a smooth product manifold. We also show how the spectrum of type
IIA theory on T$^8$/$\{1,\,{\cal I}_8\}$ can be reproduced from the
K3 $\times$ T$^4$ compactification. In the former case as also shown by Sen [9]
there
is a tadpole contribution [23,24] since the Euler charactersitic 
of the internal space
T$^8$/$\{1,\,{\cal I}_8\}$ is non-zero whereas in the latter case there is 
no tadpole since the Euler characteristic of  K3 $\times$ T$^4$ is zero.
 
Next, we consider another two dimensional orbifold model 
(K3 $\times$ T$^5$)/$\{1,\, \sigma \cdot {\cal J}_5\}$ of M-theory, where 
${\cal J}_5$ changes the sign of all the coordinates of T$^5$ and the 
three-form gauge field in M-theory. We show that this model is equivalent
to an orbifold model (K3 $\times$ T$^4$)/$\{1,\, \sigma \cdot {\cal I}_4\}$ of 
type IIB string theory, where ${\cal I}_4$ simply changes the sign of the
coordinates of T$^4$. This equivalence can also be understood, as before, 
through orbifolding 
procedure on the M-theory definition and the orbifolding procedure again 
falls into category 2
of Sen's classification. The massless spectrum [25] on the M-theory side coming 
from the untwisted sector in this case consists of a
gravity multiplet and 16 matter multiplets of the chiral N=8 or (0, 8) 
supersymmetry.
There are also 128 antichiral bosons and 64 antichiral Majorana-Weyl (M-W)
spin 1/2 fermions in the untwisted sector which remain inert under the 
supersymmetry. In the twisted sector there are 256 antichiral M-W
spin 1/2 fermions coming from the 256 fixed points of the M-theory orbifold
model as dictated by the condition of gravitational anomaly [26] cancellation.
These fermions also remain inert under supersymmetry. 
We show how this spectrum can be reproduced in the above mentioned type IIB 
orbifold model. As in the previous case, we then look at the same orbifold
(K3 $\times$ T$^4$)/$\{1,\, \sigma \cdot {\cal I}_4\}$ of type IIA string 
theory. By using the self-duality symmetry of type IIA string theory 
compactified on T$^4$ which exchanges ${\cal I}_4$ to $(-1)^{F_L}$ [6], 
this model
can be shown to be equivalent to an orbifold model T$^4$ $\times$ (K3
$\times$ S$^1$)/$\{1,\,\sigma \cdot {\cal J}_1\}$ of M-theory. We study
the spectrum of these two models and show that they match if we take into 
account the tadpole contribution of the orbifold model of type IIA theory.
This type IIA orbifold model has a gravity multiplet and 40 scalar multiplets
of the non-chiral (4, 4) supersymmetry of which 16 scalar multiplets come 
from the one-loop tadpole contribution of the Neveu-Schwarz--Neveu-Schwarz
(NS-NS) sector antisymmetric tensor field.

Sen in ref.[27] conjectured that a six dimensional orbifold model of 
M-theory on (K3 $\times$ S$^1$)/$\{1,\, {\cal J}_1\cdot \sigma\}$ is 
equivalent to an orientifold/orbifold model of type IIB theory on K3/$\{1,\,
\Omega \cdot \sigma\}$ $\equiv$ K3/$\{1,\, (-1)^{F_L} \cdot \sigma\}$ 
considered by Dabholkar and Park [28]. We refer this conjecture to be 
Dabholkar-Park-Sen (DPS) conjecture. Here $\Omega$ denotes the world-sheet
parity transformation which is a symmetry of the type IIB string theory. The 
symmetries $\Omega$ and $(-1)^{F_L}$ are conjugate to each other by the
SL(2, Z) invariance of the type IIB theory in ten dimensions. The spectrum 
in the two theories match in
an interesting way as was shown in ref.[12]. We here comment on 
how this conjecture 
can be seen to follow from Dasgupta-Mukhi-Witten [22,29] (DMW) conjecture 
about the equivalence between M-theory on T$^5$/$\{1,\,{\cal J}_5\}$ and
type IIB theory on K3 by orbifolding
procedure and assuming that it commutes with duality. In this case the 
orbifolding procedure falls into category 3 of Sen's classification, where the
equivalence between the two resulting theories is the weakest. The reason 
why we get a sensible dual pair can be traced if we start from the equivalence
of type IIA theory and M-theory on S$^1$ with the radius of the circle going
to zero i.e. from the M-theory definition. By taking orbifold on both sides
of the M-theory definition DPS conjecture can then be shown 
to follow and in this case the 
orbifolding procedure falls into category 2 in the classification.

The organization of this paper is as follows. In subsection 2.1
of section 2, we discuss the equivalence of M-theory on 
T$^9$/$\{1,\,{\cal J}_9\}$ and type IIB theory on T$^8$/$\{1,\,{\cal I}_8\}$.
In subsection 2.2, we show that the same orbifold model of type IIA theory
is equivalent to M-theory compactification on smooth product manifold 
K3 $\times$ T$^5$. We show that the spectrum in these two theories match
after we take into account the tadpole contribution on the orbifold of type IIA
side. Next, we consider the equivalence of M-theory on (K3 $\times$ T$^5$)
/$\{1, \, \sigma \cdot {\cal J}_5\}$ and type IIB theory on 
(K3 $\times$ T$^4$)/$\{1, \, \sigma \cdot {\cal I}_4\}$ in subsection 3.1 of
section 3. In subsection 3.2, we consider the same orbifold model of type IIA
theory and show that it is equivalent to M-theory compactification on the
orbifold T$^4$ $\times$ (K3 $\times$ S$^1$)/$\{1,\,\sigma \cdot {\cal J}_1\}$.
Again the spectrum matches after we take into account the tadpole contribution
on the type IIA side. Finally,
in section 4, we comment that DPS conjecture between orbifold models of type 
IIB theory and M-theory in six dimensions can be obtained from DMW conjecture
through orbifolding procedure and assuming that it commutes with duality.
Discussions and conclusions are presented in section 5.

\vspace{1 cm}

\begin{large}
\noindent{\bf 2. Orbifolds of Two Dimensional Toroidal Compactification
of M-Theory and Type II String Theories:}
\end{large}

\vspace{.5cm}

In this section, we study the orbifold models of some two dimensional
toroidal compactification of M-theory and their corresponding duals in
type II string theories. In the first part, we show that the orbifold model
T$^9$/$\{1,\,{\cal J}_9\}$ of M-theory is dual to the orbifold model
T$^8$/$\{1, \,{\cal I}_8\}$ of type IIB string theory. We compare the massless
spectrum on both sides giving evidence in favor of this conjecture. In the
second part, we show that the same orbifold model T$^8$/$\{1, \,{\cal I}_8\}$
of type IIA string theory is equivalent to a series of two dimensional 
orbifold models of M-theory and various other string theories. Interestingly,
this particular orbifold model is shown to be equivalent to type IIA 
theory compactified
on a smooth product manifold K3 $\times$ T$^4$. We compare the massless
spectra in these two models and find that they match in an interesting way.

\vspace{1cm}

\noindent{\bf 2.1 M-Theory on T$^9$/${\cal J}_9$ and Type IIB Theory on 
T$^8$/${\cal I}_8$:}

\vspace{.5cm}

The duality between the two dimensional models of M-theory and type IIB theory
in the title of this subsection has been conjectured by Dasgupta and Mukhi [22]
and also by Sen [12]. Here ${\cal J}_9 \equiv {\cal J}_1 \cdot {\cal I}_8$
is a discrete symmetry in M-theory [30] which changes the sign of all the nine 
coordinates of T$^9$ and also the sign of the three-form gauge field present
in M-theory. The spectrum for the M-theory model has been obtained in ref.[22]
by making use of the condition of the two dimensional gravitational anomaly
cancellation. We will show in this subsection how the same spectrum can be
reproduced in the type IIB model on T$^8$/$\{1,\,{\cal I}_8\}$.
We first like to
point out that this duality conjecture can be understood from the M-theory
definition i.e. we take M-theory compactified on S$^1$ as
equivalent by definition to type IIA string theory when the radius of the
circle goes to zero. We now further compactify the
theories on T$^8$ and mod out the M-theory by the symmetry group
$\{1,\,{\cal J}_1 \cdot {\cal I}_8\} \equiv \{1,\, {\cal J}_9\}$ and the
type IIA theory by the corresponding image $\{1,\,(-1)^{F_L} \cdot
{\cal I}_8\}$\footnote[2]{By looking at the massless spectrum it can be 
easily checked that the
symmetry ${\cal J}_1$ in M-theory has the same effect as $(-1)^{F_L}$ in the
type IIA theory.}. 
Since the compact space T$^9$/$\{1,\,{\cal J}_9\}$ has the
structure of S$^1$ fibered over T$^8$/$\{1,\,(-1)^{F_L} \cdot {\cal I}_8\}$,
we get the
following equivalence by applying the duality conjecture fiberwise [5]:
\eqabegin
& & {\rm M\,\,theory\,\,on\,\,\,} {\rm T}^5/\{1,\,{\cal J}_9\}\nn\\
&\equiv & {\rm Type\,\, IIA\,\,theory\,\,on\,\,\,}
{\rm T}^8/
\{1,\,(-1)^{F_L} \cdot {\cal I}_8\}
\eqaend
Note that the orbifolding procedure falls into category 2 of Sen's
classification. Now using an $R \rightarrow 1/R$ duality transformation
on one of the circles of T$^8$ we convert the type IIA model to type IIB
model on T$^8$/$\{1,\,{\cal I}_8\}$ where $(-1)^{F_L} \cdot {\cal I}_8$ in
type IIA theory gets converted\footnote[1]{This effect can also be checked
by comparing the massless spectrum in both type IIA and type IIB models.} 
to ${\cal I}_8$ in type IIB theory and thus
`proves' the duality conjecture proposed in the title of this subsection.

The massless spectrum of the M-theory model on T$^9$/$\{1,\,{\cal J}_9\}$
was shown in ref.[22] to consist of apart from a gravity multiplet, eight
scalar multiplets of (0, 16) supersymmetry of the model and 128 antichiral
bosons which remain inert under the supersymmetry. In order to cancel the
gravitational anomaly it was found that one requires 512 antichiral M-W
fermions which come from the 512 fixed points of ${\cal I}_9$ on T$^9$ of
the model. Thus summarizing the massless spectrum including the untwisted
as well as the twisted sector states we have:
\eqabegin
& &(g_{\mu\nu},\,\phi,\,16\psi_\mu^-,\,16\lambda^+)\nn\\
& &8 \times (16\phi^+,\,16\lambda^+)\\
& & 128\phi^- \qquad {\rm and}\qquad 512\lambda^-\nn
\eqaend
Note that the graviton and the gravitino have formally $-1$ degree of freedom
in two dimensions and so, the graviton has to be compensated by a scalar and
a gravitino has to be compensated by a spin 1/2 M-W fermion. 
Here we have generically denoted the scalars by $\phi$ and M-W spin 1/2
fermions by $\lambda$. Also the superscripts $(+,\,-)$ indicate, respectively,
the (chiral, antichiral) bosons and fermions. 
We will reproduce this spectrum in the type IIB theory on T$^8$/$\{1,\,
{\cal I}_8\}$.

The massless spectrum of type IIB string theory in
ten dimensions contains a graviton $g_{\mu\nu}$, an antisymmetric tensor
field $B_{\mu\nu}^{(1)}$ and a dilaton $\phi^{(1)}$ in the NS-NS sector.
In the R-R sector it contains another antisymmetric tensor field
$B_{\mu\nu}^{(2)}$, another scalar $\phi^{(2)}$ and a four-form gauge field
$A_{\mu\nu\rho\sigma}^-$ whose field strength is antiself-dual. In the NS-R
sector it has an antichiral gravitino $\psi_\mu^{(1)\,-}$ and a chiral M-W
spin 1/2 fermion $\lambda^{(1)\,+}$. In the R-NS sector also it has one
antichiral gravitino $\psi_\mu^{(2)\,-}$ and a chiral M-W spin 1/2 fermion
$\lambda^{(2)\,+}$. Let us first consider the bosonic sector. From the ten
dimensional graviton we get a graviton in two dimensions (2d) and 36 scalars. 
All the scalars
in 2d will survive the ${\cal I}_8$ projection. $B_{\mu\nu}^{(1)}$ gives 28
scalars in 2d. The dilaton $\phi^{(1)}$ and the other scalar in the R-R sector
$\phi^{(2)}$ both give one scalar each in 2d. $B_{\mu\nu}^{(2)}$ gives 28
scalars and $A_{\mu\nu\rho\sigma}^-$ gives 70 half scalars or 35 scalars in 2d.
Thus we have one graviton $g_{\mu\nu}$, one dilaton $\phi$ and $(36 + 28 + 1
+ 28 + 35) = 128$ scalars in the bosonic sector. In the fermionic sector each
of the negative chirality gravitinos of the ten dimensional type IIB theory
gives 8 gravitinos of negative chirality
and 64 positive chirality spin 1/2 M-W fermions that survive the ${\cal I}_8$
projection. Note that ${\cal I}_8$ acts on the spinors of one chirality with
$+1$ and on the other chirality with $-1$. 
Also from each of the positive chirality M-W spinors in ten
dimensions we will get 8 positive chirality M-W spinors in 2d 
that survive ${\cal I}_8$
projection. Thus in the fermionic sector we have 16 gravitinos of negative
chirality which shows that we have (0, 16) supersymmetry and $(2 \times 64 +
2 \times 8) = 144$ positive chirality M-W spinors. One can also calculate
the massless states in the twisted sector, but we notice that the condition of
gravitational anomaly cancellation dictates in this case the presence of 256
antichiral bosons. The gravitational anomaly [26] assoicated with
spin 3/2 field (chiral
minus antichiral) is $I_{3/2} = (23/24) p_1$, that of spin 1/2 field
is $I_{1/2}
= - (1/24) p_1$ and for the chiral minus antichiral boson is $I_s
= - (1/24)p_1$,
where $p_1$ is the anomaly polynomial and the fermions are the complex
fermions. So, the spectrum would be anomaly free if it satisfies $I_{3/2}
: I_{1/2} : I_s = 1 : -m : (23 + m)$, where $m$ is any integer.
The spectrum in this case satisfies $I_{3/2} : I_{1/2}
: I_s = -8 : 72 : -256 = 1 : -9 : 32 = 1 : -9 : (23+9)$ and so is anomaly
free. 
Since there are 256 fixed points of ${\cal I}_8$ on T$^8$,
so each fixed point in this orbifold model of type IIB theory will contribute
one antichiral boson. The 256 antichiral bosons can be converted to 
512 antichiral M-W fermions by bose-fermi equivalence in two dimensions 
and thus the spectrum matches precisely with the M-theory model.
Note that the orbifold model T$^8$/$\{1,\,{\cal I}_8\}$ of type IIB theory
can not be smoothed out to any known manifold like K3 $\times$ K3$'$ (since
this will break the supersymmetry by 1/4th instead of 1/2) or T$^4$ $\times$
K3 (since this will give a non-chiral theory instead of a chiral theory). So,
next we consider type IIA model on the same orbifold 
T$^8$/$\{1,\,{\cal I}_8\}$, where the situation is different.

\vspace{1cm}

\noindent{\bf 2.2 Type IIA on T$^8$/${\cal I}_8$ and M-Theory on K3 $\times$
T$^5$:}

\vspace{.5cm}

In the second part of this section, we show that an orbifold model of type 
IIA string theory on T$^8$/$\{1,\,{\cal I}_8\}$ is non-perturbatively
equivalent to M-theory model on K3 $\times$ T$^5$ which, in turn, is
equivalent to type IIA theory on K3 $\times$ T$^4$. So, we have an equivalence
between the same theory compactified on two different internal spaces, one
is an orbifold whereas, the other is a smooth product manifold. Note that
unlike in type IIB case, this equivalence is intuitively possible since type 
IIA theory compactified either on T$^8$/$\{1,\,{\cal I}_8\}$ or on 
K3 $\times$ T$^4$ gives non-chiral theories in two dimensions with (8, 8)
supersymmetry. The equivalence of these two models can be understood in several
ways. First, if we start from the M-theory definition, then,
\eqabegin
& &{\rm Type\,\,IIA\,\,theory\,\,on\,\,\,} {\rm T}^8/\{1,\,{\cal I}_8\}\nn\\
&\equiv&{\rm M\,\,theory\,\,on\,\,\,} {\rm T}^8/\{1,\,{\cal I}_8\} \times
{\rm S}^1
\eqaend
Now splitting the eight dimensional torus T$^8$ as a product of two four
dimensional torus T$^4$ $\times$ (T$^4$)$'$ as well as splitting ${\cal I}_8
\equiv {\cal I}_4 \cdot {\cal I}_4'$ and then using the self-duality symmetry
of type IIA string theory on T$^4$ which changes the geometric symmetry 
${\cal I}_4$ on T$^4$ to $(-1)^{F_L}$ we can recast the above M-theory model
to the M-theory model on (T$^4$ $\times$ (T$^4$)$'$ $\times$ S$^1$)/$\{1,\,
{\cal J}_1 \cdot {\cal I}_4'\}$. Note that we have replaced $(-1)^{F_L}$ in
type IIA theory by ${\cal J}_1$ in M-theory. Thus we arrive at the equivalence:
\eqabegin
& &{\rm Type\,\,IIA\,\,theory\,\,on\,\,\,} {\rm T}^8/\{1,\,{\cal I}_8\}\nn\\
&\equiv&{\rm M\,\,theory\,\,on\,\,\,} {\rm T}^4 \times 
{\rm T}^5/\{1,\,{\cal J}_5\} 
\eqaend
The M-theory model in eq.(4) is equivalent, by DMW conjecture, to type IIB 
theory on T$^4$ $\times$ K3. By T-duality this type IIB model is also 
equivalent to type IIA theory on T$^4$ $\times$ K3 which, in turn, is 
equivalent to M-theory compactified on K3 $\times$ T$^5$. Thus we have the
following chain of equivalences between various models:
\eqabegin
& &{\rm Type\,\,IIA\,\,theory\,\,on\,\,\,} {\rm T}^8/\{1,\,{\cal I}_8\}\nn\\
&\equiv&{\rm M\,\,theory\,\,on\,\,\,} {\rm T}^4 \times {\rm T}^5/\{1,\,
{\cal J}_5\}\nn\\
&\equiv& {\rm Type\,\,IIB\,\,theory\,\,on\,\,\,} {\rm T}^4 \times {\rm K3}\\
&\equiv&{\rm Type\,\,IIA\,\,theory\,\,on\,\,\,} {\rm T}^4 \times {\rm K3}\nn\\
&\equiv&{\rm M\,\,theory\,\,on\,\,\,} {\rm K3} \times{\rm T}^5\nn
\eqaend

We can also understand the equivalence of the models mentioned in the title
of this subsection if we start from the equivalence of
type IIB model on T$^8$/$\{1,\,(-1)^{F_L} \cdot {\cal I}_8\}$ and type IIB
model on T$^8$/$\{1,\,\Omega \cdot {\cal I}_8\}$, where $\Omega$ is the
world-sheet parity invariance of the type IIB theory. These models were shown
to be equivalent by Sen [9] as $(-1)^{F_L}$ gets precisely
converted to $\Omega$
by the SL(2, Z) invariance of the type IIB string theory in ten dimensions. The
first theory is equivalent to type IIA theory on T$^8$/$\{1,\,{\cal I}_8\}$
and the second theory is equivalent by $R \rightarrow 1/R$ duality
transformation on all
the circles of T$^8$ to type IIB theory on T$^8$/$\{1,\,\Omega\}$, which is
nothing but type I theory on T$^8$. By the ten dimensional string-string
duality of type I and heterotic string theory with gauge group
SO(32), we get the second
model to be equivalent to heterotic string theory on T$^8$. By the six
dimensional
string-string duality between heterotic string on T$^4$ and type IIA theory
on K3, we get the model to be equivalent to type IIA theory on K3 $\times$
T$^4$ which, in turn, is equivalent to M-theory on K3 $\times$ T$^5$. We thus
have the following chain of dualities:
\eqabegin
& &{\rm Type\,\,IIA\,\,theory\,\,on\,\,\,} {\rm T}^8/\{1,\,{\cal I}_8\}\nn\\
&\equiv& {\rm Type\,\,I\,\,theory\,\,on\,\,\,} {\rm T}^8\nn\\
&\equiv&{\rm Heterotic\,\,string\,\,theory\,\,on\,\,\,} {\rm T}^8\\
&\equiv&{\rm Type\,\,IIA\,\,theory\,\,on\,\,\,} {\rm K3} \times {\rm T}^4\nn\\
&\equiv&{\rm M\,\,theory\,\,on\,\,\,} {\rm K3} \times{\rm T}^5\nn
\eqaend

We now show that the massless spectra in the two different type IIA models
on ${\rm T}^8/\{1,\,{\cal I}_8\}$ and ${\rm K3} \times {\rm T}^4$, indeed, are
the same.
The massless spectrum of type IIA string theory in ten dimensions has a
graviton $g_{\mu\nu}$, an antisymmetric tensor field $B_{\mu\nu}$ and
a dilaton $\phi$ in the NS-NS sector while in the R-R sector
it has a gauge field $A_\mu$ and a three-form antisymmetric tensor field
$A_{\mu\nu\rho}$. In the fermionic sector it contains a positive chirality
gravitino $\psi_\mu^+$ and a positive chirality M-W spinor $\lambda^+$
in the NS-R sector, whereas in the R-NS sector it has a negative chirality
gravitino $\psi_\mu^-$ and a negative chirality M-W spinor $\lambda^-$.

We first consider the orbifold model of type IIA string theory on 
${\rm T}^8/\{1,\,{\cal I}_8\}$. This model has already been studied by Sen
in ref.[9]. We will study this model in more detail and from a different point
of view. In the bosonic sector of this type IIA reduction, we
will get in 2d one graviton $g_{\mu\nu}$ and 36 scalars from the 
ten dimensional
graviton, we also get 28 scalars from $B_{\mu\nu}$ and one scalar from the
dilaton. Note that both $A_{\mu}$ and $A_{\mu\nu\rho}$ will not give any
scalar in 2d since they change sign under ${\cal I}_8$. 
Thus we have one graviton,
one dilaton and $(36 + 28) = 64$ scalars in the bosonic sector. In the
fermionic sector, we will get from the ten dimensional gravitino
$\psi_\mu^+$, 8 gravitinos of positive
chirality $\psi_\mu^+$ in 2d and 64 negative chirality M-W spinors $\lambda^-$
that survive the ${\cal I}_8$ projection. Also from ten dimensional M-W
spinor $\lambda^+$ we get 8 positive chirality M-W spinors in 2d. In the R-NS
sector we get 8 gravitinos of negative chirality and 64 positive chirality
M-W spinors in 2d from the ten dimensional gravitino $\psi_\mu^-$.
Finally, from
$\lambda^-$ in ten dimensions we get 8 negative chirality M-W spinors in 2d. 
Thus collecting all the states in the untwisted sector we have one graviton
$g_{\mu\nu}$, one dilaton $\phi$, 64 scalars and 8 gravitinos of positive
chirality $\psi_\mu^+$, 8 gravitinos of negative chirality $\psi_\mu^-$
also 72 positive chirality $\lambda^+$ and 72 negative chirality $\lambda^-$
M-W spinors:
\eqabegin
& & (g_{\mu\nu},\,\phi,\,8\psi_\mu^+,\,8\lambda^-,\,8\psi_\mu^-,\,8\lambda^+)
\nn\\
& &8\times(8\phi^+,\,8\lambda^+),\qquad {\rm and}\qquad
8\times(8\phi^-,\,8\lambda^-)
\eqaend
It can be easily checked that there are no massless states in the twisted
sector of this theory. The massless states in the twisted sector can only
arise in the R-R sector since the left or right moving fermions in the NS
sector gives vacuum energy greater than zero. But the R-R sector ground state
in this case does not survive GSO projection [9].

In studying the consequences of the six-dimensional string-string duality
between type IIA string theory on K3 and heterotic string theory on T$^4$,
it was found by Vafa and Witten [23] that certain two dimensional
compactifications of string theories are inconsistent because of the presence
of tadpoles of the NS-NS sector antisymmetric
tensor field. In these computations of tadpoles, it was realized later by
Sethi, Vafa and Witten [24] that the tadpole contribution
is simply proportional
to the Euler characteristic $\chi$ of the compact eight dimensional manifold
divided by 24. It was, therefore, argued that the two dimensional
compactification of string theories can be made consistent by introducing
$\chi$/24 number of one-branes in the internal space. So, if $\chi$ is not
divisible by 24 or is negative\footnote[2]{When $\chi$ is negative one
requires anti-branes to cancel the tadpoles. Since anti-branes carry wrong
chirality matter multiplets this breaks the supersymmetry
of the model but the compactification could be consistent [24].}, then
the tadpoles
can not be removed and the corresponding two-dimensional compactification
remains inconsistent.

The two dimensional type IIA compactification we are discussing also contains
one-loop tadpole from the NS-NS sector antisymmetric tensor field. The tadpole
contribution in this theory can be easily calculated by computing the Euler
characteristic of the eight dimensional compact space T$^8$/$\{1,\,
{\cal I}_8\}$. The result is\footnote[1]{This should be understood as the 
Euler characteristic of the smooth manifold formed after blowing up 
the orbifold as usual.} [31]: 
\begineq
\chi(\frac{{\rm T}^8}{{\cal I}_8}) = (0 - 256)/2 + 2 \times 256 = 384
\endeq
where $\chi({\rm T}^8) = 0$, 256 is the number of fixed points in this model
and 2 is the order of the symmetry group. Thus we find that the tadpoles can
be cancelled by introducing 384/24 = 16 elementary type IIA strings in the
internal space as noted also by Sen [9] using different method. The collective
coordinates of the 16 type IIA strings with (8, 8) supersymmetry will give
128 bosons, 128 positive chirality and 128 negative chirality fermions
as additional massless states. Adding these states in (7), we find the
complete massless spectrum of type IIA
theory on T$^8$/$\{1,\,{\cal I}_8\}$ to be:
\eqabegin
& & (g_{\mu\nu},\,\phi,\,8\psi_\mu^+,\,8\lambda^-,\,8\psi_\mu^-,\,8\lambda^+)
\nn\\
& &24\times(8\phi^+,\,8\lambda^+),\qquad {\rm and}\qquad
24\times(8\phi^-,\,8\lambda^-)
\eqaend
Thus this model has (8, 8) supersymmetry and 24 scalar multiplets alongwith
a gravity multiplet.

We will see that the same spectrum can be reproduced by compactification of
type IIA theory on K3 $\times$ T$^4$. Under K3 reduction $g_{\mu\nu}$ gives
one graviton and 58 scalars in six dimensions; these, in turn, under T$^4$
reduction gives one graviton, 10 scalars and 58 scalars (from 58 scalars
in 6d) in 2d. Similarly, $B_{\mu\nu}$ gives 28 scalars, 6 from T$^4$ reduction
and 22 from K3 reduction. The dilaton gives one scalar in 2d. The gauge field
$A_\mu$ gives 4 scalars in 2d (it does not give any scalar under K3 reduction).
Finally, the three-form gauge field gives one three-form gauge field and 22
vector gauge fields under K3 reduction. Under T$^4$ reduction, $A_{\mu\nu\rho}$
gives 4 scalars and the 22 vector gauge fields give 88 scalars in 2d. Thus
in the bosonic sector of the K3 $\times$ T$^4$ reduction of type IIA theory
we have $(g_{\mu\nu},\,\phi,\,192\,\, {\rm scalars})$.

In the fermionic sector the ten dimensional positive chirality gravitino
gives one gravitino of positive chirality and 20 Weyl spinors of negative
chirality in six dimensions under K3 reduction. They under T$^4$ reduction
gives 4 gravitinos of positive and 4 gravitinos of negative chirality. It
also gives 32 M-W spinors --- 16 each with both positive and negative 
chiralities. Then from 20
Weyl spinors of negative chirality we get 80 M-W spinors of positive and 80
M-W spinors of negative chirality. Also from a single positive
chirality M-W spinor in ten dimensions we get 4 of positive chirality and
4 of negative chirality M-W spinors in 2d. In the R-NS sector the spectrum
is exactly the same. So, collecting all the states in the fermionic sector
we have $2 \times (16 + 80 + 4) = 200$ positive chirality and 
200 negative chirality M-W spinors apart
from 8 gravitinos of both positive and negative chiralities. 
This is precisely the spectrum we have obtained from the type IIA 
compactification on
T$^8$/$\{1,\,{\cal I}_8\}$. We also like to point out that there is no tadpole
contribution of this type IIA model on K3 $\times$ T$^4$ since the Euler
characteristic of this product manifold is zero. Note that the tree-level
spectrum of type IIA theory on K3 $\times$ T$^4$ is identical with the 
tree-level and one-loop level spectrum of type IIA theory on 
T$^8$/$\{1,\,{\cal I}_8\}$, indicating that the two models are 
non-perturbatively equivalent. For a perturbative T-duality we expect the 
spectrum to match separately at every perturbative level.

\vspace{1cm}           
   
\begin{large}
\noindent{\bf 3. Orbifolds of Two Dimensional Compactification
of M-Theory and Type II String Theories Involving K3:}
\end{large}

\vspace{.5cm}

In this section, we study some orbifold models of two dimensional 
compactification of M-theory and type II string theories involving K3. In the
first part of this section, we consider an orbifold compactification of
M-theory on (K3 $\times$ T$^5$)/$\{1,\,\sigma \cdot {\cal J}_5\}$ and show
that it is equivalent to the orbifold compactification of type IIB string 
theory on (K3 $\times$ T$^4$)/$\{1,\,\sigma \cdot {\cal I}_4\}$. By comparing
the massless spectra of both these models we provide evidence in favor of
this duality conjecture. In the second part, as in the previous section, we 
show that the same orbifold model of type IIA theory is dual to an orbifold
compactification of M-theory on T$^4$ $\times$ (K3 $\times$ S$^1$)/$\{1,\,
\sigma \cdot {\cal J}_1\}$. We compute the massless spectra of these two
models and show that they match if we include the tadpole contribution on the
type IIA side.

\vspace{1cm}

\noindent{\bf 3.1 M-Theory on (K3 $\times$ T$^5$)/$\sigma \cdot {\cal J}_5$
and Type IIB Theory on (K3 $\times$ T$^4$)/$\sigma \cdot {\cal I}_4$:}

\vspace{.5cm}

In this subsection, we study a two dimensional orbifold model of M-theory
involving K3 and find the corresponding dual model in the type IIB theory.
This
provides another dual pair between M-theory and type IIB theory and will be
shown to give consistent anomaly-free spectrum having chiral (0, 8)
supersymmetry. The duality conjecture mentioned in the title of this 
subsection
can again be understood from the M-theory definition. So, as in the previous
section, we start with the equivalence of
M-theory on S$^1$ and type IIA theory when the radius of the circle goes to
zero. We then further compactify the models
on K3 $\times$ T$^4$ and mod out by the combined group of discrete
transformation $\sigma \cdot {\cal J}_5\, \equiv\,\sigma \cdot {\cal J}_1 
\cdot {\cal I}_4$ in the M-theory side and the corresponding image 
$\sigma \cdot (-1)^{F_L}
\cdot {\cal I}_4$ on the type IIA side. Since (K3 $\times$ T$^5$)
/$\{1,\, \sigma \cdot {\cal J}_5\}$ has the structure S$^1$ fibered over
(K3 $\times$ T$^4$)/$\{1,\, \sigma \cdot (-1)^{F_L} \cdot {\cal I}_4\}$, by
applying duality conjecture fiberwise [5] we get the equivalence:
\eqabegin
& & {\rm M\,\,theory\,\,on\,\,\,} ({\rm K3} \times {\rm T}^5)/\{1,\,
\sigma \cdot{\cal J}_5\}\nn\\
&\equiv & {\rm Type\,\, IIA\,\,theory\,\,on\,\,\,}
({\rm K3} \times {\rm T}^4)/
\{1,\,\sigma \cdot (-1)^{F_L} \cdot {\cal I}_4\}
\eqaend
Here $\sigma$ denotes an involution
on K3 surface. This involution has been used [32] to construct 
the type IIA dual
of the maximally supersymmetric six dimensional heterotic string
compactification of Chaudhuri, Hockney and Lykken [33]. By
constructing a special K3
surface it has been shown in [32] that $\sigma$ changes the sign of the
8 harmonic
(1, 1) forms of the K3 surface leaving the other 12 harmonic (1, 1) forms as
well as the (0, 0), (0, 2), (2, 0) and (2, 2) forms invariant as a consequence
of the Lefschetz fixed point theorem. Thus $\sigma$ acts on K3 with eight
fixed points and exchanges the two E$_8$ factors in the lattice of the second
cohomology elements of K3. We also note that $\sigma$ does not break the
supersymmetry of the model unlike another involution of K3 considered in the
literature [2] known as Enriques involution which does not
leave the holomorphic
(2, 0) form invariant and breaks the supersymmetry of the model by half.
Now, by making an $R \rightarrow 1/R$ duality transformation on one 
of the circles
of T$^4$ on the type IIA side in eq.(10) we convert this 
model to type IIB model on
(K3 $\times$ T$^4$)/$\{1,\, \sigma \cdot {\cal I}_4\}$, where as before
$(-1)^{F_L} \cdot {\cal I}_4$ in type IIA theory gets converted to ${\cal I}_4$
in type IIB theory. Thus we `derived' the proposed duality conjecture between
the M-theory and the type IIB model from the M-theory definition. Note again
that the orbifolding procedure falls into category 2 of Sen's classification.

In order to compare the massless spectrum of these two models, we mention that
that the spectrum for the M-theory on (K3 $\times$ T$^5$)
/$\{1,\, {\cal J}_5 \cdot \sigma\}$ has already been obtained by Kumar and Ray
in ref.[25]. It has been found that this model contains a gravity multiplet,
16 scalar multiplets of (0, 8) supersymmetry and also 128 antichiral bosons
as well as 64 antichiral M-W spin 1/2 fermions as massless states in the
untwisted sector where the latter states remain inert under supersymmetry. 
In order to cancel the two dimensional gravitational anomaly
it was found that the twisted sector of this theory contributes 256 antichiral
M-W spin 1/2 fermions from the $32 \times 8 = 256$ fixed points. These 
twisted sector states also remain inert under supersymmetry. So
summarizing the spectrum we have,
\eqabegin
& &(g_{\mu\nu},\,\phi,\,8\psi_\mu^-,\,8\lambda^+)\nn\\
& & 16 \times (8\phi^+,\,8\lambda^+)\nn\\
& &(128\phi^-,\,64\lambda^-)\\
& & 256\lambda^-\nn
\eqaend
We now see how the same spectrum can be obtained from the type IIB model on
(K3 $\times$ T$^4$)/$\{1,\, \sigma \cdot {\cal I}_4\}$.
Since various K3 reductions of type IIB string theory has been performed in 
detail in ref.[34], we will be brief here. 

We first consider the bosonic sector of type IIB string theory.
The massless spectrum for the ten dimensional type IIB string theory is given
in the previous section. The ten dimensional graviton will give
one graviton in 2d, 10 scalars from T$^4$ reduction which
remain
invariant under ${\cal I}_4$ and 34 scalars from K3 reduction which remain
invariant
under $\sigma$. From $B_{\mu\nu}^{(1)}$ we get six scalars 
from T$^4$ and 14 scalars
from K3 since 14 out of 22 two-forms remain invariant under $\sigma$. From the
dilaton $\phi^{(1)}$, we get one scalar in 2d. In the R-R sector, we get one
scalar from $\phi^{(2)}$. We also get 20 scalars from $B_{\mu\nu}^{(2)}$, 
6 from T$^4$ reduction and 14 from K3 reduction. Finally, from
$A_{\mu\nu\rho\sigma}^-$, we get one scalar in 6d for K3 reduction half of
which comes from dualizing $A_{\mu\nu\rho\sigma}^-$ corresponding to (0, 0)
form on K3 and another half comes from (2, 2) form on K3. This gives one
scalar in 2d. We also get from it 11 self-dual two-forms and 3 anti self-dual
two-forms corresponding to 11 anti self-dual two-forms and 3 self-dual
two-forms on K3 that remain invariant under $\sigma$. These two-forms, in turn,
will give $(6 \times 11 + 6 \times 3) = 84$ half-scalars or 42 scalars under
T$^4$ reduction. Thus altogether we have, one graviton, one dilaton and
$(10 + 34 + 6 + 14 + 1 + 6 + 14 + 1 + 42) = 128$ scalars in the bosonic sector.

In the fermionic sector, we get from $\psi_\mu^{(1)\,-}$ one gravitino of
negative chirality on K3 reduction (this is a Weyl spinor in 6d) which, 
in turn,
gives 4 gravitinos of negative chirality and 16 positive chirality M-W spin
1/2 fermions in 2d after ${\cal I}_4$ projection. It also gives 48 positive
chirality M-W spinors in 2d corresponding to 12 positive chirality Weyl spinors
which remain invariant under $\sigma$. Finally, we get 32 negative chirality
M-W spinors in 2d corresponding to 8 positive chirality
Weyl spinors in 6d which
change sign under $\sigma$. From $\lambda^{(1)\,+}$ we will get 4 positive
chirality M-W spinors in 2d. In the R-NS sector $\psi_\mu^{(2)\,-}$ and
$\lambda^{(2)\,+}$ give the identical states. So, collecting all the states
in the fermionic sector we get 8 gravitinos of negative chirality $\psi_\mu^-$,
$2 \times (16 + 48 + 4) = 136$ positive chirality M-W spinors $\lambda^+$
and $2 \times 32 = 64$ negative chirality M-W spinors $\lambda^-$.

Thus the complete spectrum can be arranged as a gravity multiplet $(g_{\mu\nu},
\,\phi,\,8\psi_\mu^-,\,8\lambda^+)$, 16 scalar multiplets $(8\phi^+,\,
8\lambda^+)$ and $(128\phi^-,\,64\lambda^-)$ which remain inert under (0, 8)
supersymmetry of the model. This is the identical spectrum for the 
corresponding M-theory model in the untwisted sector. 
It can be easily checked that this spectrum is anomalous. In this case
we find that for the spectrum to become anomaly free we need 128 antichiral
bosons. The spectrum then satisfies $I_{3/2} : I_{1/2} : I_s = -4 : 36 : -128
= 1: -9 : (23+9)$ and is anomaly free. Since we have precisely
128 fixed points in this orbifold
model ($\sigma$ has eight fixed points on K3 and ${\cal I}_4$ has sixteen
fixed points on T$^4$), so, each fixed point will contribute an antichiral
boson in the spectrum. The twisted sector of the type IIB model on
(K3 $\times$ T$^4$)/$\{1,\,\sigma \cdot {\cal I}_4\}$ has been
obtained by Sen in
ref.[9] and it was found to contain 128 antichiral bosons as we found from the
anomaly cancellation. These 128 antichiral bosons can be converted to 256
negative chirality M-W spin 1/2 fermions by bose-fermi equivalence and thus
we find that the spectrum precisely matches with that of the M-theory model
as obtained by Kumar and Ray [25].
                                                                     
It is clear in this case that there are no tadpoles in the two dimensional
M-theory model on (K3 $\times$ T$^5$)/$\{1,\,\sigma \cdot {\cal J}_5\}$ or
equivalently the type IIA model on (K3 $\times$ T$^4$)/
$\{1,\,\sigma \cdot (-1)^{F_L} \cdot {\cal I}_4\}$. Note here that $(-1)^{F_L}$
does not pose any problem in the calculation since this type
IIA model is equivalent to type IIB model on (K3 $\times$ T$^4$)/
$\{1,\, \sigma \cdot {\cal I}_4\}$ by an $R \rightarrow 1/R$ duality
transformation on one of the circles of T$^4$. Since the type IIB model does
not contain $(-1)^{F_L}$ factor, we can use the world-sheet parity invariance
underwhich the NS-NS sector antisymmetric tensor field $B_{\mu\nu}^{(1)}$ 
changes sign and thus the tadpole contribution vanishes [23].
The situation is completely different if we consider the same orbifold 
compactification, namely, (K3 $\times$ T$^4$)/$\{1,\,\sigma \cdot {\cal I}_4\}$
of type IIA theory to which we turn next.

\vspace{1cm}

\noindent{\bf 3.2 Type IIA Theory on 
(K3 $\times$ T$^4$)/$\sigma \cdot {\cal I}_4$
and M-Theory on T$^4$ $\times$ (K3 $\times$ S$^1$)/$\sigma \cdot {\cal J}_1$:}

\vspace{.5cm}

The equivalence between the two models mentioned in the title of this 
subsection can be understood from the M-theory definition if we use the 
self-duality symmetry [6] of type IIA string theory compactified on T$^4$. So,
we start from the equivalence of M-theory on S$^1$ and type IIA string theory 
when the radius of S$^1$ goes to zero. If we then further compactify both
the theories on K3 $\times$ T$^4$ and mod out by the symmetry $\sigma \cdot
{\cal I}_4$ on both sides, we have the following equivalence:
\eqabegin
& & {\rm M\,\,theory\,\,on\,\,\,} ({\rm K3} \times {\rm T}^4)/\{1,\,
\sigma \cdot{\cal I}_4\} \times {\rm S}^1\nn\\
&\equiv & {\rm Type\,\, IIA\,\,theory\,\,on\,\,\,}
({\rm K3} \times {\rm T}^4)/\{1,\,\sigma \cdot {\cal I}_4\}
\eqaend
Now using the self-duality symmetry of type IIA theory on T$^4$ which exchanges
${\cal I}_4$ to $(-1)^{F_L}$ we can convert the above M-theory model to 
M-theory on $({\rm K3} \times {\rm T}^4)/\{1,\,\sigma \cdot (-1)^{F_L}\}
\times {\rm S}^1$ which is nothing but M-theory on $({\rm K3} \times {\rm T}^4
\times {\rm S}^1)/\{1,\,\sigma \cdot {\cal J}_1\}$. Since the orbifold group
$\{1,\,\sigma \cdot {\cal J}_1\}$ does not act on T$^4$, we can extract it 
out to write
\eqabegin
& & {\rm M\,\,theory\,\,on\,\,\,} {\rm T}^4 \times ({\rm K3} 
\times {\rm S}^1)/\{1,\,\sigma \cdot {\cal J}_1\}\nn\\
&\equiv & {\rm Type\,\, IIA\,\,theory\,\,on\,\,\,}
({\rm K3} \times {\rm T}^4)/\{1,\,\sigma \cdot {\cal I}_4\}
\eqaend
`proving' the proposed duality conjecture. Note that this M-theory model is
equivalent, by DPS conjecture, to type IIB theory on ${\rm T}^4 \times
{\rm K3}/\{1,\,\sigma \cdot (-1)^{F_L}\} \equiv {\rm T}^4 \times
{\rm K3}/\{1,\,\sigma \cdot \Omega\}$.

We now compute the massless spectra of both the models. Let us first look at
the M-theory orbifold model on ${\rm T}^4 \times ({\rm K3}
\times {\rm S}^1)/\{1,\,\sigma \cdot {\cal J}_1\}$. The six dimensional 
orbifold model of M-theory on $({\rm K3}\times {\rm S}^1)/\{1,\,\sigma 
\cdot {\cal J}_1\}$ has already been studied by Sen in ref.[27]. It has been
found that this model contains apart from a gravity multiplet $(g_{\mu\nu},\,
\psi_\mu^{\alpha},\, A_{\mu\nu}^-)$, 8 vector multiplets $(A_\mu,\, \lambda^
{\alpha})$, one tensor multiplet $(A_{\mu\nu}^+,\,\lambda_{\alpha},\,\phi)$
and 12 hypermultiplets $(\lambda_{\alpha},\,4\phi)$ of N=1, D=6 supersymmetry
algebra in the untwisted sector. In the twisted sector this models contains
8 M-theory five-branes which support 8 tensor and 8 hypermultiplets and the 
complete spectrum including the untwisted as well as the twisted sector states
have been shown to be free of gravitational anomaly. Here, $A_{\mu\nu}^-(
A_{\mu\nu}^+)$ denotes the antiself-dual(self-dual) antisymmetric tensor. 
$\alpha$ is the spinor index for which up(down) indicates antichiral(chiral)
Weyl spinors. So, we just have to consider the T$^4$ reduction. In the 
bosonic sector, we will get one graviton and 10 scalars in 2d from the six
dimensional graviton $g_{\mu\nu}$. From $A_{\mu\nu}^-$ we will get 6 
half-scalars or 3 scalars in 2d. 8$A_\mu$ from the 8 vector multiplets will
give 32 scalars and 9$A_{\mu\nu}^+$ from the 9 tensor multiplets will give 54
half-scalars or 27 scalars in 2d. Also, 9 scalars from 9 tensor multiplets
in six dimensions will give 9 scalars in 2d and 20 $\times$ 4 = 80 scalars from
20 hypermultiplets will give 80 scalars in 2d. So, altogether in the bosonic 
sector we have one $g_{\mu\nu}$, one dilaton $\phi$ and $(9 + 3 + 32 + 27 +
9 + 80) = 160$ scalars after T$^4$ reduction.

Similarly, in the fermionic sector a six dimensional negative chirality Weyl
gravitino will give 4 negative and 4 positive chirality M-W gravitinos in 2d.
It will also give 16 positive and 16 negative chirality M-W spin 1/2 fermions
in 2d. From 8 negative chirality Weyl spinors of 8 vector multiplets, we will
get 32 negative and 32 positive chirality M-W spinors in 2d. Also, from 9 
positive chirality Weyl spinors of 9 tensor multiplets in 6d, we will get 36
positive chirality and 36 negative chirality M-W spinors in 2d after T$^4$
reduction. Finally, from the 20 positive chirality Weyl spinors of 20
hypermultiplets in 6d, we will get 80 positive and 80 negative chirality M-W
spin 1/2 fermions in 2d. So, collecting all the states in the fermionic sector
we have 4 M-W gravitinos of positive and negative chiralities and $(16 + 32
+ 36 + 80) = 164$ M-W spin 1/2 fermions of both chiralities. We can arrange 
the complete spectrum in a gravity multiplet and 40 scalar multiplets of (4, 4)
supesymmetry in 2d as follows,
\eqabegin
& &(g_{\mu\nu},\,\phi,\,4\psi_\mu^+,\,4\lambda^-,
\,4\psi_\mu^-,\,\lambda^+)\nn\\
& & 40 \times (4\phi^+,\,4\lambda^+) \qquad {\rm and} \qquad 
40 \times (4\phi^-,\,4\lambda^-)
\eqaend   
We will see in the following that an identical massless spectrum can be 
reproduced in the two dimensional orbifold model of type IIA theory on
$({\rm K3} \times {\rm T}^4)/\{1,\,\sigma \cdot {\cal I}_4\}$. The massless 
spectrum of type IIA string theory in ten dimensions has been given in section
2.2. We here consider the reduction. The ten dimensional graviton will give
a six dimensional graviton and 34 scalars after K3 reduction which remain
invariant under $\sigma$. They, in turn, will give one graviton and 10 scalars
for T$^4$ reduction which remain invariant under ${\cal I}_4$. We will also
get 34 scalars in 2d from 34 scalars in 6d. The ten dimensional antisymmetric
tensor $B_{\mu\nu}$ will give one antisymmetric tensor and 14 scalars for
K3 reduction which remain invariant under $\sigma$. In 2d they will give 20
scalars, six of which come from T$^4$ reduction of $B_{\mu\nu}$ in 6d. The
ten dimensional dilaton will give one scalar in 2d. In the R-R sector the
vector gauge field will not give any scalar in 2d since it changes sign under
${\cal I}_4$. From the three-form gauge field we will get, in six dimensions,
one three-form gauge field, 14 vector gauge fields which remain invariant
under $\sigma$ and 8 vector gauge fields which change sign under $\sigma$.
The three-from gauge field and 14 vector gauge fields will not give any scalar
for T$^4$ reduction since they change sign under ${\cal I}_4$. But, from 8
vector gauge fields which change sign for K3 reduction under $\sigma$ will
give 32 scalars for T$^4$ reduction that again change sign under ${\cal I}_4$.
So, in the bosonic sector we have altogether one graviton, one dilaton and
$(10 + 34 + 6 + 14 + 32) = 96$ scalars in 2d.

Let us next consider the fermionic sector. Type IIA theory in ten dimensions
has a positive chirality M-W gravitino and a positive chirality M-W spin 1/2
fermion in the NS-R sector. In the R-NS sector it has a negative chirality
gravitino and a negative chirality spin 1/2 M-W fermion. For K3 reduction, the
ten dimensional $\psi_\mu^+$ will give one positive chirality Weyl gravitino,
12 negative chirality Weyl spinors that remain invariant under $\sigma$ and
8 negative chirality Weyl spinors that change sign under $\sigma$. The six
dimensional Weyl gravitino, in turn, for T$^4$ reduction will give 4 positive
chirality M-W gravitinos and 16 negative chirality M-W spin 1/2 fermions
that remain invariant under ${\cal I}_4$. From the 12 negative chirality Weyl
spinors, we will get 48 negative chirality M-W spin 1/2 fermions for T$^4$
reduction that remain invariant under ${\cal I}_4$. Also, from 8 negative
chirality Weyl spinors which changed sign under $\sigma$, we will get 32
positive chirality M-W spin 1/2 fermions under T$^4$ reduction which remain
invariant under ${\cal I}_4$. Finally, the ten dimensional positive chirality
M-W spinor will give 4 positive chirality M-W spin 1/2 fermions in 2d. In the
R-NS sector after the reduction, we will get the 
identical fermionic states with
opposite chiralities. So, altogether we have 4 gravitinos of positive 
and 4 gravitinos of negative chiralities. We also have $(16 + 48 + 32 + 4) 
= 100$ positive and 100 negative chirality M-W spin 1/2 fermions. Therefore,
the states in the untwisted sector can be arranged as a gravity multiplet and
24 scalar multiplets as follows:
\eqabegin
& &(g_{\mu\nu},\,\phi,\,4\psi_\mu^+,\,4\lambda^-,
\,4\psi_\mu^-,\,\lambda^+)\nn\\
& & 24 \times (4\phi^+,\,4\lambda^+) \qquad {\rm and} \qquad
24 \times (4\phi^-,\,4\lambda^-)
\eqaend
Comparing this spectrum with (14), we find that we still need 16 more scalar
multiplets in the type IIA orbifold model on $({\rm K3} \times {\rm T}^4)
/\{1,\, \sigma \cdot {\cal I}_4\}$ to match the spectrum with the corresponding
M-theory model. It can be easily checked again, as in the type IIA theory
on ${\rm T}^8/\{1,\,{\cal I}_8\}$, that there are no twisted sector states in
this type IIA model since the massless states in the twisted sector can only
arise as R-R ground state. But, the R-R ground state in this type IIA model
does not survive GSO projection. The additional states, as we will see,  
come from the one-loop tadpole contribution of the type IIA orbifold model
on $({\rm K3} \times {\rm T}^4)/\{1,\, \sigma \cdot {\cal I}_4\}$.
The tadpole contribution can be easily calculated, as before, by calculating
the Euler characteristic of the compact eight dimensional internal space.
The Euler characteristic of the smooth manifold obtained by blowing up the
orbifold $({\rm K3} \times {\rm T}^4)/\{1,\, \sigma \cdot {\cal I}_4\}$ is
given as,
\eqabegin
\chi(\frac{{\rm K3} \times {\rm T}^4}{{\cal I}_4 \cdot \sigma}) &=&
(0 - 128)/2 + 2 \times 128\nn\\
&=& 192
\eqaend
where the Euler characteristic of K3 $\times$ T$^4$ is zero, 128 is the number
of fixed points of the orbifold model ($\sigma$ has 8 fixed points on K3 and
${\cal I}_4$ has 16 on T$^4$) and 2 is the order of the symmetry group.
Since 192 is divisible by 24, we can make the model consistent by introducing
192/24 = 8 elementary type IIA strings in the internal space. This result can
also be understood as follows. Note that locally the space $({\rm K3} \times 
{\rm T}^4)/\sigma \cdot {\cal I}_4$ has the structure of T$^8$/${\cal I}_8$
and  therefore, the physics in the neighborhood of the fixed points should
remain the same in both the models.
We have mentioned in section 2.2 that the tadpoles in the type IIA model
on T$^8$/$\{1,\,{\cal I}_8\}$ can be removed by introducing 16 elementary
type IIA strings of (8, 8) supersymmetry. So, each fixed point in these type
IIA models acts as a source of $-(1/16)$ unit of antisymmetric tensor field
charge. Since in the type IIA model on T$^8$/$\{1,\,{\cal I}_8\}$ there are 
256 fixed points and each string carries a single unit of antisymmetric tensor
field charge, the charge can be canceled by introducing 16 strings in the
internal space. Now, the orbifold model of type IIA theory we are discussing
here has 128 fixed points containing total 128$\times (-1/16) = - 8$ units
of antisymmetric tensor field charge which can be canceled by introducing
8 elementary type IIA strings of (8, 8) supersymmetry in the internal space.
The collective coordinates of 8 elementary type IIA strings with (8, 8)
supersymmetry will give rise to 64 scalars and 64 positive as well as 64 
negative chirality M-W spin 1/2 fermions. They can be arranged as 16 matter
multiplets of (4, 4) supersymmetry of the model we are considering. Now,
adding these states in (15), we find that the complete massless spectrum of the
type IIA orbifold model on   
$({\rm K3} \times {\rm T}^4)/\{1,\, \sigma \cdot {\cal I}_4\}$ is given as,
\eqabegin
& &(g_{\mu\nu},\,\phi,\,4\psi_\mu^+,\,4\lambda^-,
\,4\psi_\mu^-,\,\lambda^+)\nn\\
& & 40 \times (4\phi^+,\,4\lambda^+) \qquad {\rm and} \qquad
40 \times (4\phi^-,\,4\lambda^-)
\eqaend 
This is precisely the spectrum we have obtained for the orbifold 
compactification of M-theory on T$^4$ $\times$ (K3 $\times$ S$^1$)/$\{1,\,
\sigma \cdot {\cal J}_1\}$ and thus provides evidence in favor of the duality
conjecture proposed in the title of this subsection.

\vspace{1cm}

\begin{large}
\noindent{\bf 4. DPS Conjecture from DMW Conjecture:}
\end{large}

\vspace{.5cm}

It has been conjectured by Dasgupta and Mukhi [22] and also independently by
Witten [29] that a six dimensional orbifold compactification of M-theory on
T$^5$/$\{1,\,{\cal J}_5\}$ is equivalent to a six dimensional orbifold
compactification of type IIB theory on T$^4$/$\{1,\,{\cal I}_4\} \simeq
$ K3. In a different work Sen [27] has conjectured that a six 
dimensional orbifold
model of M-theory on (K3 $\times$ S$^1$)/$\{1,\,{\cal J}_1 \cdot \sigma\}$
is equivalent to a six dimensional orientifold model studied by Dabholkar
and Park [28] of type IIB theory on K3/$\{1,\,\Omega \cdot \sigma\}$. This 
orientifold model by ten dimensional SL(2, Z) invariance is equivalent to
six dimensional orbifold model of type IIB theory on K3/$\{1,\,(-1)^{F_L} 
\cdot \sigma\}$. It is easy to see that both these conjectures follow from
the M-theory definition. In fact, it has already been pointed out by Sen [21]
how DMW conjecture follows from the M-theory definition. In this section, we 
will first show how DPS conjecture follows also from the M-theory definition
and then point out that it can be obtained from DMW conjecture as well through
orbifolding procedure and assuming that the orbifolding procedure in this
case commutes with duality.

By definition M-theory compactified on S$^1$ is equivalent to type IIA theory.
We then further compactify both theories on K3 and mod out M-theory by the
combined group of discrete symmetries $\{1,\,{\cal J}_1 \cdot \sigma\}$ and
type IIA theory by the corresponding image $\{1,\,(-1)^{F_L} \cdot \sigma\}$.
Since the space (K3 $\times$ S$^1$)/$\{1,\,{\cal J}_1 \cdot \sigma\}$ has
the structure S$^1$ fibered over K3/$\{1,\,(-1)^{F_L} \cdot \sigma\}$, by
applying duality conjecture fiberwise we get the following equivalence:
\eqabegin
& &{\rm M\,\,theory\,\,on\,\,\,} ({\rm K3} \times {\rm S}^1)/\{1,\,{\cal J}_1
\cdot \sigma\}\nn\\
&\equiv & {\rm Type\,\,IIA\,\,theory\,\,on\,\,\,} {\rm K3}/\{1,\,(-1)^{F_L}
\cdot \sigma\}
\eqaend
Now by going to the orbifold limit of K3 we write the right hand side of 
(18) as type IIA theory on T$^4$/$\{1,\,{\cal I}_4,\,(-1)^{F_L} \cdot
\eta \cdot {\cal I}_4,\,(-1)^{F_L} \cdot \eta\}$, where T$^4$ denotes the 
product of four circles with coordinates $(X^6,\,X^7,\,X^8,\,X^9)$ at the
self-dual radius and $\eta$ denotes the operation $(X^6,\,X^7,\,X^8,\,X^9)
\rightarrow (X^6,\,X^7,\,X^8,\,X^9 + \pi R_9)$, with $R_9$ denoting the
radius of the ninth circle. Also the involution
$\sigma$ on K3 surface is represented as $\eta \cdot {\cal I}_4$ on T$^4$.
Note that although $\eta \cdot {\cal I}_4$ has 16 fixed points on T$^4$,
the orbifold identifies them pairwise leaving 8 fixed points as $\sigma$
on K3.
It has been shown by Sen in ref.[12] that this type IIA model is equivalent
by an $R \rightarrow 1/R$ duality transformation on one of the circles
other than the ninth circle to type IIB
model on the same orbifold i.e. T$^4$/$\{1,\,(-1)^{F_L} \cdot \eta \cdot 
{\cal I}_4,
\,{\cal I}_4,\,(-1)^{F_L} \cdot \eta\} \simeq {\rm K3}/\{1,\,
(-1)^{F_L} \cdot \sigma\}$. The only difference is that the elements of the
orbifold group got reshuffled in the type IIB theory. Thus it shows that DPS
conjecture indeed follows from the M-theory definition. Note that in this case
the orbifolding procedure falls into the category 2 of Sen's classification.

Now we start from DMW conjecture i.e.
\eqabegin
& &{\rm M\,\,theory\,\,on\,\,\,} {\rm T}^5/\{1,\,{\cal J}_5\}\nn\\
&\equiv& {\rm Type\,\,IIB\,\,theory\,\,on\,\,\,} {\rm K3} \simeq 
{\rm T}^4/\{1,\,{\cal I}_4\}
\eqaend
Then we mod out the type IIB theory on K3 by the symmetry group 
$\{1,\,(-1)^{F_L} \cdot \sigma\}$ i.e. by $\{1,\,(-1)^{F_L} \cdot \eta \cdot
{\cal I}_4\}$ on T$^4$/$\{1,\,{\cal I}_4\}$ and M-theory by the corresponding 
image $\{1,\,\eta \cdot {\cal I}_4\}$. If we assume that the orbifolding
procedure commutes with duality in this case we arrive at the following
equivalence
\eqabegin
& &{\rm M\,\,theory\,\,on\,\,\,} {\rm T}^5/\{1,\,{\cal J}_5,\,\eta \cdot 
{\cal I}_4,\,{\cal J}_1 \cdot \eta\}\nn\\
&\equiv& {\rm Type\,\,IIB\,\,theory\,\,on\,\,\,} 
{\rm T}^4/\{1,\,{\cal I}_4,\,(-1)^{F_L} \cdot \eta \cdot {\cal I}_4,
\,(-1)^{F_L}
\cdot \eta\}
\eqaend 
We now trade in the coordinates $X^m$ of T$^5$ of M-theory in favor of new 
coordinates $Z^m$ as
\begineq
Z^m = X^m \qquad {\rm for} \qquad 6\leq m \leq 8, \qquad Z^9 = X^9 + \pi R_9/2,
\qquad Z^{10} = X^{10}
\endeq
then in terms of the new coordinates we have the left hand side of (20) as 
M-theory on (T$^4$ $\times$ S$^1$)
/$\{1,\,{\cal J}_5 \cdot \eta,\, {\cal I}_4,\,{\cal J}_1 \cdot \eta\}
\simeq$ (K3 $\times$ S$^1$)/$\{1,\,{\cal J}_1 \cdot \sigma\}$. Thus we have
arrived at DPS conjecture starting form DMW conjecture.

Note that in `deriving' DPS conjecture this way we have directly taken 
the orbifold 
projection without further compactifying the theories on another manifold.
So, for this case the orbifolding procedure falls into category 3 of Sen's
classification, where it was mentioned that the duality conjecture for the
resulting theories is the weakest. In fact, in most of the cases this way of 
orbifolding does not lead to sensible dual pairs. The reason why it works in
this case is because the same duality conjecture, as we have seen, can be 
obtained from the M-theory definition when the orbifolding procedure falls
into category 2.

\vspace{1cm}

\begin{large}
\noindent{\bf 5. Discussion and Conclusion:}
\end{large}

\vspace{.5cm}

We have studied in this paper several examples of dual pairs involving 
M-theory and type II string theories in two space-time dimensions. 
Dual pairs are 
obtained from the M-theory definition and through orbifolding procedure
of category 2 of the classification made by Sen. In particular, we have 
considered two dimensional orbifolds of both toroidal compactification
and compactification involving a single K3 of M-theory and type II string
theories. By analyzing the massless spectrum we have provided evidence in
favor of the duality conjecture between M-theory on T$^9$/$\{1,\,{\cal J}_9\}$
and type IIB theory on T$^8$/$\{1,\,{\cal I}_8\}$. This provides an example
of consistent anomaly-free chiral supergravity theory in two dimensions having
(0, 16) supersymmetry. We then pointed out that the same orbifold model 
T$^8$/$\{1,\,{\cal I}_8\}$ of type IIA theory is inconsistent because of the
presence of tadpoles of the NS-NS sector antisymmetric tensor field. This
inconsistency can be removed by introducing 16 elementary type IIA strings
in the internal space. We have shown that this type IIA orbifold model is
dual to M-theory compactified on a smooth product manifold K3 $\times$ T$^5$
or equivalently to type IIA theory on K3 $\times$ T$^4$. The analysis of the 
massless spectrum is much simpler in the latter case since this is a smooth
manifold having Euler characteristic zero and we just had to calculate the
untwisted sector states. We have shown that the spectrum matches with the
type IIA theory on T$^8$/$\{1,\,{\cal I}_8\}$ if we take into account 
the additional states required to remove the inconsistency involving tadpoles.
Next, we have considered another two dimensional orbifold model involving a
single K3. In this case, again by analyzing the spectrum on both sides, we
have provided evidence in favor of the duality conjecture between M-theory
on (K3 $\times$ T$^5$)/$\{1,\,\sigma \cdot {\cal J}_5\}$ and type IIB theory
on (K3 $\times$ T$^4$)/$\{1,\,\sigma \cdot {\cal I}_4\}$. In this case, we
have an example of anomaly-free chiral supegravity theory in two dimensions
having (0, 8) supersymmetry. Although this type IIB orbifold model does not 
have tadpoles, the same orbifold model of type IIA theory is inconsistent
because of the presence of tadpoles. We have shown that the orbifold model
(K3 $\times$ T$^4$)/$\{1,\,\sigma \cdot {\cal I}_4\}$ of type IIA theory is
dual to the orbifold model T$^4$ $\times$ (K3 $\times$ S$^1$)
/$\{1,\,\sigma \cdot {\cal J}_1\}$ of M-theory. We have shown that the 
massless spectrum in these two theories match if we take into account the 
additional states required to remove the inconsistency involving tadpoles on
the type IIA side. Finally, we have pointed out that the DPS conjecture for 
the six dimensional orbifold models involving type IIB theory and M-theory
simply follows from DMW conjecture of the equivalence of M-theory on 
T$^5$/$\{1,\,{\cal J}_5\}$ and type IIB theory on K3. 
Our analysis confirms that when the
orbifolding procedure falls into category 2 of the classification made by Sen,
it leads to sensible dual pairs of the resulting theories even if the
original pairs involve M-theory on one side.

We have not considered
in this paper the two dimensional orbifold compactifications involving two
K3's. For example, it can be easily seen that the orbifold models of M-theory
on (K3 $\times$ K3$'$ $\times$ S$^1$)/$\{1,\, \sigma \cdot \sigma' \cdot 
{\cal J}_1\}$ is dual to type IIB theory on (K3 $\times$ T$^4$)
/$\{1,\,(-1)^{F_L} \cdot {\cal I}_4\} \cdot \{1,\,\sigma \cdot \eta \cdot 
{\cal I}_4\}$, where $\eta$ is as given in section 4. This duality can also be
understood from M-theory definition and through orbifolding procedure of
category 2. Similarly, it can be seen again from orbifolding procedure of
category 2 on the M-theory definition that the orbifold model 
(K3 $\times$ K3$'$)/$\{1,\,(-1)^{F_L}\cdot \sigma \cdot \sigma'\}$ of type IIB
string theory is dual to the orbifold model (K3 $\times$ T$^5$)
/$\{1, {\cal J}_5\} \cdot \{1,\,\sigma \cdot \eta \cdot {\cal I}_4\}$ of 
M-theory. Note that in both cases, the orbifold group in the dual model has
the structure {\bf Z}$_2$ $\times$ {\bf Z}$_2$ and can not be further
simplified to a single {\bf Z}$_2$. It has been recognized in ref.[17] that
for orbifolding group greater than {\bf Z}$_2$, there are some ambiguities in
the analysis of the spectrum and requires a careful study for establishing the
duality conjecture. We leave this problem for a separate investigation.

\vspace{1cm}

\begin{large}

\noindent{\bf Acknowledgements:}

\end{large}

\vspace{.5cm}

I would like to thank Ashoke Sen for very helpful discussions. This work is 
supported in part by the Spanish Ministry of Education (MEC) fellowship.

\vspace{1cm}

\begin{large}

\noindent{\bf References:}
\end{large}

\vspace{.5cm}

\begin{enumerate}
\item S. Ferrara, J. Harvey, A. Strominger and C. Vafa, \pl 361 (1995) 59.
\item J. A. Harvey, D. A. Lowe and A. Strominger, \pl 362 (1995) 65.
\item S. Kachru and C. Vafa, \np 450 (1995) 69.
\item A. Klemm, W. Lerche and P. Mayr, \pl 357 (1995) 313.
\item C. Vafa and E. Witten, {\it Dual String Pairs with N=1 and N=2 
Supersymmetry in Four Dimensions}, hep-th/9507050.
\item A. Sen and C. Vafa, \np 455 (1995) 165. 
\item G. Aldazabal, A. Font, L. Ibanez and F. Quevedo, \np 461 (1996) 537.
\item B. Hunt and R. Schimmrigk, \pl 381 (1996) 427. 
\item A. Sen, \np 474 (1996) 361.
\item B. Hunt, M. Lynker and
R. Schimmrigk, {\it Heterotic/Type II Duality in D=4 and String-String
Duality}, hep-th/9609082.
\item M. Duff, R. Minasian and E. Witten, \np 465 (1995) 413.
\item A. Sen, Mod. Phys. Lett A11 (1996) 1339.
\item B. S. Acharya, {\it N=1 Heterotic/M-Theory Duality and Joyce Manifolds},
hep-th/9603033; {\it M-Theory Compactification and Two-Brane/Five-Brane
Duality}, hep-th/9605047.
\item D. Morrison and C. Vafa, \np 473 (1996) 74; {\it Compactifications
of F-Theory on Calabi-Yau Threefolds II}, hep-th/9603161.  
\item A. Sen, {\it F-Theory and Orientifolds}, hep-th/9605150.
\item J. Polchinski, {\it Tensors from K3 Orientifolds}, hep-th/9606165.
\item R. Gopakumar and S. Mukhi, {\it Orbifold and Orientifold 
Compactifications of F-Theory and M-Theory to Six and Four Dimensions},
hep-th/9607057.
\item J. Blum and A. Zaffaroni, {\it An Orientifold from F-Theory}, 
hep-th/9607019.
\item A. Dabholkar and J. Park, {\it A Note on Orientifolds and F-Theory},
hep-th/9607041.
\item J. Blum, {\it F-Theory Orientifolds, M-Theory Orientifolds and 
Twisted Strings}, hep-th/9608053.
\item A. Sen, {\it Unification of String Dualities}, hep-th/9609176. 
\item K. Dasgupta and S. Mukhi, \np 465 (1996) 399.
\item C. Vafa and E. Witten, \np 447 (1995) 261.
\item S. Sethi, C. Vafa and E. Witten, {\it Constraints on Low Dimensional
String Compactifications}, hep-th/9606122.
\item A. Kumar and K. Ray, {\it Compactification of M-Theory to Two
Dimensions}, hep-th/9604164.
\item L. Alvarez-Gaume and E. Witten, \np 234 (1984) 269.
\item A. Sen, Phys. Rev. D53 (1996) 6725.
\item A. Dabholkar and J. Park, {\it An Orientifold of Type IIB Theory on K3},
hep-th/9602030; {\it Strings on Orientifolds}, hep-th/9604178.
\item E. Witten, \np 463 (1996) 383.
\item P. Horava and E. Witten, \np 460 (1996) 506.
\item M. B. Green, J. H. Schwarz and E. Witten, {\it Superstring Theory},
Vol. II, Cambridge University Press, 1987.
\item J. Schwarz and A. Sen, \pl 357 (1995) 323.
\item S. Chaudhuri, G. Hockney and J. Lykken, {\it Maximally Supersymmetric
String Theories in D $<$ 10}, hep-th/9505054.
\item S. Roy, {\it An Orbifold and an Orientifold of Type IIB Theory on
K3 $\times$ K3}, hep-th/9607157 (to appear in Phys. Lett. B.)

\end{enumerate}

\vfil
\eject 

\end{document}